# The Application of Distributed Control Algorithms using VOLTTRON-based Software Platform

Jingwei Luo, *Student Member*, *IEEE*
Hajir Pourbabak, *Student Member*, *IEEE*
Department of Electrical and Computer Engineering
University of Michigan-Dearborn
Dearborn, MI, USA

Wencong Su, *Member*, *IEEE*
Department of Electrical and Computer Engineering
University of Michigan-Dearborn
Dearborn, MI, USA
Email: wencong@umich.edu

*Abstract*—This paper gives an insight into the applications of an open-source control system platform named VOLTTRON. This platform was developed by the Pacific Northwest National Laboratory. A brief introduction is given on the functionality and key features of the platform. Potential applications in the areas of building control and electric vehicle charging are stated, along with an overview of existing projects. A comparison is also made between VOLTTRON and other related software. An actual implementation case of VOLTTRON is then presented in the case study. The demonstration uses the VOLTTRON platform as a message bus. Decentralized generators and consumers are simulated by 16 single-board computers.

## I. Introduction

The power grid is undergoing a transformation from a monopolized control system to a more decentralized one. Many new distributed issues such as renewable energy generation, responsive loads and automation in the distribution system are posing a new challenge to the traditional centralized control method [1]. A new group of agent-based software platforms is emerging to cope with this new challenge. Software agents can make local decisions based on information provided by other agents or online sources. This capability suits the decentralized control demand of the future smart grid.

VOLTTRON[TM] is an open-source control system platform developed by the U.S. Department of Energy (DOE) at the Pacific Northwest National Laboratory (PNNL). The platform provides services such as resource management, agent code verification and directory services [1]. These services allow the platform to manage different assets within the power system. In the large scale, VOLTTRON[TM] can manage assets within smart grids, achieve demand response matching, assist with energy trading and record grid data. In a smaller scale, VOLTTRON[TM] can be a good platform for the operation and optimization of large commercial buildings.

VOLTTRON[TM] has the potential to play an important role in both the commercial and academic realm [2]. VOLTTRON[TM] allows researchers to quickly test new control methods on a multi-agent environment and can serve as an ideal platform for rapid-prototyping of new control applications. In addition, the open-source nature of the platform frees developers from any license issues. Such openness also reflects VOLTTRON[TM]'s ability to merge with other platforms and software. Moreover, due to its agent mobility service, VOLTTRON[TM] agents can be transferred from platform to platform.

The concept of an agent is defined as a computer system capable of autonomous action in a situated environment [3]. An agent can react to information sent by the operator, the environment or other agents in real time. VOLTTRON[TM]'s agents are generally divided into three classes [1]. Platform agents serve as service providers in the platform. Cloud agents serve as bridges between the platform and other remote platforms. Control agents control the actual hardware devices.

Fig. 1 shows the basic components of the VOLTTRON[TM] platform. The message bus in the center allows agents in the platform to exchange information by publishing and subscribing to topics. The mechanism of the "publish" and "subscribe" paradigm will be discussed later in the case study section. Drivers are connected to the control objects and their controllers. Drivers, loggers and an archiver send a copy of the data to the historian in the cloud for backup.

This paper provides a summary of the applications of VOLTTRON[TM] in multiple aspects of a smart grid. In section II, the advantages of the VOLTTRON[TM] platform and some related projects are discussed. In section III, VOLTTRON[TM] is compared with two agent-based software platforms and one similar building control platform. In sections IV and V, the problem formulation and the case study show a detailed implementation of VOLTTRON™ in a smart grid simulation.

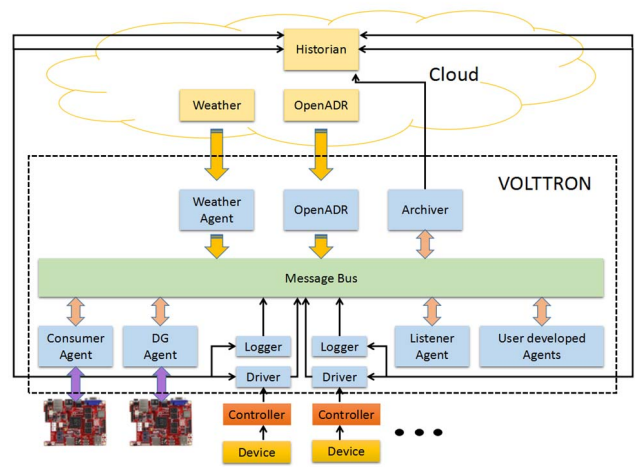

Fig. 1. Basic principle of VOLTTRON

## II. APPLICATION OF VOLTTRON

This section discusses different applications of the VOLTTRON™ platform. The two main application areas of VOLTTRON™ are building control and electric vehicle charging.

### A. Applications in Buildings

Worldwide, buildings cover a major part of energy consumption and greenhouse gas (GHG) emissions. Up to 40% of the total energy in the United States is consumed by buildings, which also constitute 38% of the GHG emissions of the country [2]. Considering the fact that the future smart electricity market will most likely have dynamic pricing, an implementation of up-to-date sophisticated building automation systems (BASs) with VOLTTRON™ could have a large impact on reducing energy consumption, GHG emissions and operation costs of buildings.

A distributed agent-based control platform is a natural fix for building control: 1) VOLTTRON™ allows different kinds of easy access. Different communication protocols, such as BACnet, Modbus, and OpenADR, are supported by the platform, making it an ideal platform to combine the information of different assets with multiple communication protocols. 2) VOLTTRON™ gives timely feedback and runs a real-time control cycle. 3) Agents can be easily installed and uninstalled on the platform in a manner similar to that used with app stores. This allows the corresponding devices to join and leave the platform easily. 4) A VOLTTRON™ system can be implemented in a low-cost credit-card-sized single board computer such as Raspberry Pie or Cubieboard, making it easy to implement.

The software nature of the agents allows VOLTTRON™ developers to program them to do a variety of tasks. Any complex control algorithm can be easily implemented into a control agent. A scalable supervisory control algorithm for the heating, ventilation, and air-conditioning (HVAC) system in large buildings using generalized gossip and the VOLTTRON™ platform is proposed by Chinde [4]. They divided a large-scale building into several different zones and used a distributed optimization method to reduce energy consumption. They also developed an optimization agent in VOLTTRON™ to run the simulation. VOLTTRON™ then served as both a control center and a message bus, with devices connected to VOLTTRON™ through BACnet.

VOLTTRON™ is a command-line-based system suitable for developers, as it gives developers greater freedom to modify the agents and achieve a wide range of functions. But being command-line-based also makes it unfriendly to users with little programming experience. A building management system with a graphical user interface (GUI) would have a much greater advantage than a system without one. Thankfully, VOLTTRON™'s open-source nature also makes it an ideal software framework to build upon.

Developers at Virginia Tech built a Building Energy Management Open Source Software (BEMOSS) platform based on the VOLTTRON™ platform [5]. Their platform aims at the optimized control of small- and medium-sized buildings and can control loads in buildings and implement demand response to save operation costs. The platform consists of multiple layers, with VOLTTRON™ being the core component of the operating system and framework layer. One of the highlights of their system is the user interface layer. The user can choose between different comfort levels associated with different costs.

### B. Applications in Electric Vehicles and Smart Grids

As the world's fossil fuel starts to run out, in the near future, electric vehicles (EV) will start to replace conventional vehicles and play a greater role in transportation. Most designs of the currently running utility grid do not consider large amounts of electric vehicles charging at the same time. Using VOLTTRON™ as an agent-based distributed control would have serval advantages: 1) The control is conducted locally, thus reducing the complexity for the utility to conduct long-distance central control. 2) Distributed control is more robust, meaning a single point of failure won't imperil the whole system. 3) The real-time nature of the system ensures a fast response to local condition changes.

The potential of VOLTTRON™'s application in integrating electric vehicles and the smart grid is discussed in [1]. They built a demo with a smart home that has three EV chargers. Three types of agents were developed within the platform to share and broadcast the information, determine charging priority of the EVs within the local region and provide a user interface to record the charging data. Such a system determines when to charge an EV based on three main factors: 1) The expected time of departure 2) The charge needed and 3) Regional information, such as the power available. The order of the priority is also determined by these factors through different agents of different households. This VOLTTRON™-based system not only reduces the cost for EV owners to charge their vehicles, but also provides information to the utility and helps them manage load and demand more easily.

As the percentage of EV ownership increases in densely populated urban areas, large EV charging decks will start to appear. These large charging facilities can be created by converting existing parking structures. Imagine a large-scale implementation of VOLTTRON™ and its impact on EV charging. The platform can provide distributed control over larger parking decks. Utilities and parking deck owners can also use VOLTTRON™ to solve demand response problems and adjust their prices accordingly.

VOLTTRON™ can also be used to solve demand response problems in smart grids. The generation of energy is becoming more decentralized with more and more renewable energy generation devices being installed in homes all over the Unitized State. The electricity market of the future will be a dynamically priced free market. The traditional centralized control method has become unsuitable to control such a fast-changing system, meaning the decentralized control method and system will play a greater role from now on. In particular, the VOLTTRON™ system can be a low-cost solution for local neighborhoods to regulate and control renewable generation devices.

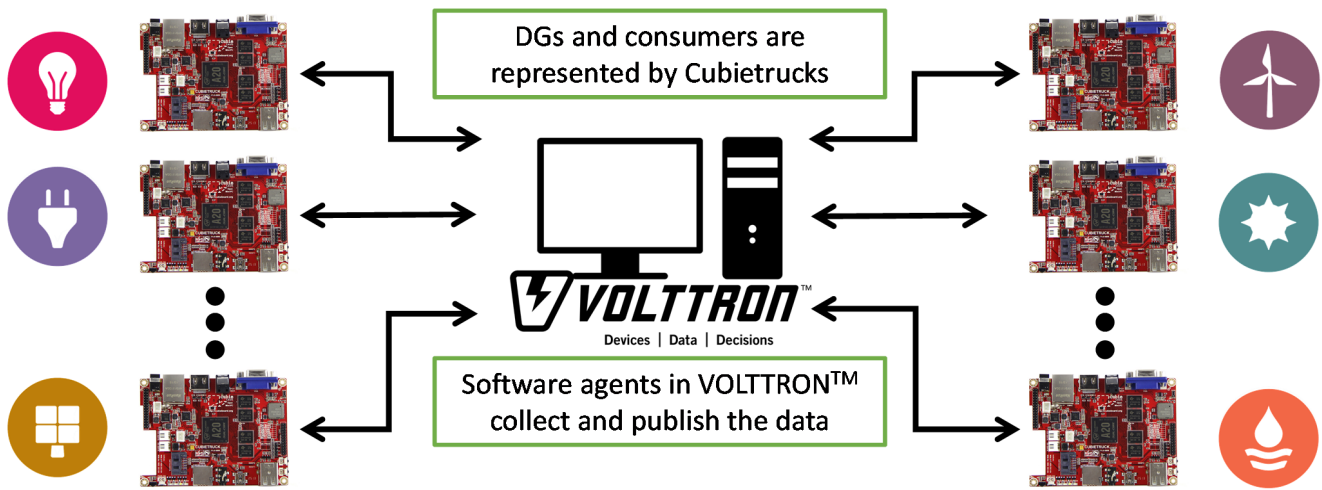

Fig. 2. An overview of the implementation

## III. RELATED SOFTWARE

Many other multi-agent platforms and decentralized control software are also available. This section gives a brief introduction to these systems and compares them with VOLTTRON$^{TM}$. The Java Agent Development framework (JADE) is a distributed middleware system. It has a flexible infrastructure which allows easy add-on of new modules [6]. The communication model of JADE is implemented based on FIPA specifications and follows a peer-to-peer paradigm [7]. Each player in the smart grid is represented by a container, and each container can host multiple agents. One platform will have one main container and many peripheral containers. The main container hosts two special agents which facilitate agent communication [8]. The peripheral containers are instances of a runtime environment in JADE. The JADE platform and its agents are coded in Java. In addition, JADE has a graphical user interface (GUI) for debugging, which makes it more user-friendly than VOLTTRON$^{TM}$. It also has an interface with Matlab Simulink, which allows developers to directly test their Simulink model in an agent-based platform.

ZEUS is a developer tool for building collaborative multi-agent applications. Its agent establishing technology allows users to rapidly develop multi-agent systems. The ZEUS toolkit consists of three basic component groups: an agent component library, agent building software and utility agents [9]. Each agent is made of multiple basic parts that allow it to exchange and handle messages, schedule reactions and store data. Like JADE, ZEUS is also a Java-based system and is FIPA compliant. The key advantage of ZEUS is that it modularizes the functions of an agent, so that a developer can simply choose the functions needed from the classes in the agent component library. ZEUS has a GUI and a runtime environment for programming and debugging agents, making it an ideal user-friendly tool for starters. One limitation of the ZEUS platform, however, is the lack of detailed documentation [3].

Mortar.io is an open-source platform for building automation. It provides services and applications to monitor and control different assets in a building [10]. Similar to VOLTTRON$^{TM}$, Mortar.io uses a publish/subscribe architecture for message exchange; however, the pub/sub architecture for Mortar.io is different from that of VOLTTRON$^{TM}$. Instead of publishing and subscribing to topics as agents do in VOLTTRON$^{TM}$, Mortar.io's pub/sub architecture publishes or subscribes to an address or a node. The Mortar.io system consists of multiple layers, each of which provides a unique function from communicating with a control object to exchanging messages between agents to providing service and a user interface.

Compared to the software mentioned above, VOLTTRON$^{TM}$ has its own advantages and drawbacks. Unlike JADE and ZEUS, which are generalized agent platforms designed to be used in multiple domains, VOLTTRON$^{TM}$ is specifically designed for the smart grid. VOLTTRON$^{TM}$'s smart-grid-facing design makes it easily compliable with the existing communication protocol in buildings and smart grid control. Its suitability for use with multiple protocols also gives VOLTTRON$^{TM}$ the advantage of much better hardware driver support [3]. JADE and ZEUS's agents are restricted to being programmed in Java. Although most of VOLTTRON$^{TM}$'s agents are written in Python, VOLTTRON$^{TM}$ does allow agents to be written in other languages. The mobility service even allows agents from other platforms to be transferred into VOLTTRON$^{TM}$. Compared with Mortar.io, a software specifically designed for building automation, VOLTTRON$^{TM}$ has a wider range of applications in the smart grid. VOLTTRON$^{TM}$'s agent platform nature also makes it more flexible and more suitable for developers.

That said, VOLTTRON$^{TM}$ has some shorting comings as well. Its lack of a GUI makes developing VOLTTRON$^{TM}$ agents harder for beginners. Additionally, VOLTTRON$^{TM}$ can only be used in power industry applications.

## IV. PROBLEM FORMULATION

This section gives a brief introduction to the consensus-based distributed algorithm proposed in [11]. Details regarding the modeling of the consumers and DGs are provided. The multiple constraints of the formulation are also discussed. The algorithm aims at solving the economic dispatch (ED) problem and maximizes the welfare for both the consumers and the distributed generators (DGs).

The ED problem is a short-term resource management problem in which the DGs should satisfy the load requirement of the smart grid in an optimized cost-effective way. This section formulates the ED problem with the definition of the consumers' utility function and the DGs' cost function.

*A. Utility and Cost Functions*

Equation (1) denotes the utility functions of the consumers. Each consumer's behavior is reflected as the parameters $\sigma$ \$/kWh$^2$ and $\omega$ \$/kWh.

$$U_j(P_j^{Load}) = \begin{cases} \omega_j P_j^{load} - \sigma_j(P_j^{load})^2 & P_j^{load} \leq \omega_j/2\sigma_j \\ \frac{\omega_j^2}{4\sigma_j} & P_j^{load} \geq \omega_j/2\sigma_j \end{cases} \quad (1)$$

The DGs are modeled by a quadratic mathematical equation. Each DG is customized by its unique set of parameters of $\alpha$ \$/kWh$^2$, $\beta$ \$/kWh and $\gamma$ \$/h. $P_i^{Gen}$ denotes the power generated by DG number i.

$$C_i(P_i^{Gen}) = \alpha_i(P_i^{Gen})^2 + \beta_i P_i^{Gen} + \gamma_i, \; i \in S_G \quad (2)$$

The objective function maximizes the summation of the utility functions (1) and minimizes the summation of the cost functions (2). Therefore, it aims to maximize the profits for both the consumers and the DGs. The objective function is shown below:

$$\text{Min}\left(\sum_{i \in S_G} C_i(P_i^{Gen}) - \sum_{j \in S_D} U_j(P_j^{load})\right) \quad (3)$$

Grid constraints such as power losses and transmission capacity are not considered in this model. One constraint that is considered, however, is the power balance between consumption $\sum_{j \in S_D}(P_j^{load})$ and aggregated generations $\sum_{i \in S_G}(P_i^{Gen})$.

$$\sum_{j \in S_D} P_j^{load} = \sum_{i \in S_G} P_i^{Gen} \quad (4)$$

The constraints of the maximum capacities of consumers and DGs are also considered. These two constraints are as shown below:

$$\begin{matrix} 0 \leq P_i^{Gen} \leq P_{i,max}^{Gen} & i \in S_G \\ 0 \leq P_j^{load} \leq P_{j,max}^{load} & i \in S_D \end{matrix} \quad (5)$$

*B. Distributed Algorithm for Economic Dispatch*

When it comes to optimization problems such as economic dispatch, a consensus-based distributed algorithm can provide a great optimized solution. In this approach, the optimized consensus can be reached by information exchange between agents within a local network or platform. A consensus is reached if and only if the value of the states of all agents are equal [12][13]. The consensus reached is based on a common agreement between all the agents within the platform. Equation (6) is the Laplacian potential for a graph. It represents the virtual energy stored in a graph [14].

$$L_P = \sum_{i,j} a_{i,j}(x_j - x_i) = 2x^T L x \quad (6)$$

This virtual energy stored in a graph shows the level of total disagreement among all agents in the platform. A consensus can only be reached by agent communication. If network-wide communication is not available, an agent should at least be able to communicate with its neighboring agents. A general consensus means there is no disagreement among all the agents. Therefore, in this optimal solution we have $L_P=0$ or $x_j=x_i$.

The consensus in this paper is considered a zero-power-mismatch consensus. The total power mismatch in the system is considered the kind of virtual energy mentioned above. According to the definition of Laplacian potential, this virtual energy must be minimized in order to reach consensus. The consensus condition is shown in (7).

$$\Delta P_1 = \Delta P_2 = \ldots = \Delta P_n = 0 \quad (7)$$

This equation means the power mismatch of the entire system estimated by every DG in the system equals zero. The agents' single-integrator dynamic property is considered and is defined as a standard linear consensus protocol (8).

$$\dot{x}_i(t) = \sum_j a_{ij}(x_j - x_i) \quad (8)$$

V. CASE STUDY

In this case study, the VOLTTRON™ message bus is implemented on a Linux desktop PC with an FX-4100 CPU @ 3.60GHz, 8-GB RAM memory. The nodes in the system are simulated by a cluster of low-cost credit-card-sized single board PCs (Cubietruck) with an All Winner Tech SOC A20 processor, ARM® Cortex™-A7 Dual-Core.

This case study is a demo for a consensus-based distributed algorithm proposed in [11]. The algorithm uses a consensus-based approach to solve the ED problem of distributed generators. The consensus-based method allows local agents to iteratively exchange information through two-way communication channels. Each agent adjusts its output accordingly with the information it receives. In the end, a global optimal decision can be reached. This section focuses on the details of the technical approach of the simulation. The simulation consists of 16 nodes, 10 of which are consumers and 6 of which are DGs. Each node is represented by a Cubietruck.

*A. The VOLTTRON™ Message Bus*

The VOLTTRON™ software is installed on a desktop PC with a Linux operating system. By following the VOLTTRON™ user guide [15], the prerequisite software was installed and the system is set up.

Two VOLTTRON™ software agents are designed to collect and forward the information from the Cubietrucks. The consumer software agent collects information from the 10 Cubietrucks representing the 10 consumer nodes. Similarly, the DG software agent collects information from the six DG nodes. Both software agents receive the data through universal asynchronous receiver/transmitter (UART) serial communication with the Cubietrucks. The software agents in VOLTTRON™ are designed to scan the serial ports from the data, with the scan and publish operation designed to run in a loop. The design references the concept of the heartbeat of the "listener agent."

TABLE I. THE EXPERIMENTAL TEST RESULTS

| Node | L1 | L2 | L3 | L4 | L5 | L6 | L7 | L8 | L9 | L10 | DG1 | DG2 | DG3 | DG4 | DG5 | DG6 |
|---|---|---|---|---|---|---|---|---|---|---|---|---|---|---|---|---|
| Output (kW) | 52.4 | 58.9 | 54.9 | 0.0 | 32.2 | 40.8 | 71.8 | 39.1 | 40.6 | 30.2 | 106.6 | 124.4 | 0.0 | 89.8 | 0.0 | 100.2 |
| Total Generation (kW) | | | | Total Load (kW) | | | | Lambda ($/kWh) | | | Iteration Number | | | | | |
| 421.0 | | | | 421.1 | | | | 7.37 | | | 42 | | | | | |

The "listener agent" is one of the sample agents in the VOLTTRON[TM] user guide [15]. The listener listens (subscribes) to all messages published in the message bus and publishes a copy of these messages. The listener is a good tool for debugging in the message bus. It can also be used as a base platform to develop more complex agents. The heartbeat of the listener is the time gap between two scan and publish operations. It is implemented by a publish heartbeat function within the software agents. The new agents used in this paper use a similar heartbeat function with a much faster heart rate. The reason for a much faster heartbeat rate is that it allows our consumer and DG software agent to collect data much faster and achieve real-time simulation. The received information is then converted to the VOLTTRON[TM] message format and published in the message exchange bus. Both agents are written in Python with the guiding of the VOLTTRON[TM] user guide [15].

The received information is then converted to the VOLTTRON[TM] message format and published in the message exchange bus. Both agents are written in Python with the guiding of the VOLTTRON[TM] user guide [15].

The information is exchanged in the VOLTTRON[TM] message bus following the "publish" and "subscribe" paradigm. Each VOLTTRON[TM] message has a topic. A topic can have multiple subtopics, following the format "topic/subtopic/subtopic/…." For example, the consumer software agents publish the data from consumer one following the topic format of "Data/consumer /consumer1." Consequently, a DG who is interested in getting only the information from consumer one can subscribe to the topic of "Data/consumer/consumer1." If the same DG is also interested in the information other consumers send, it can subscribe to a higher-level subtopic, "Data/consumer," to receive all the data sent by the consumers.

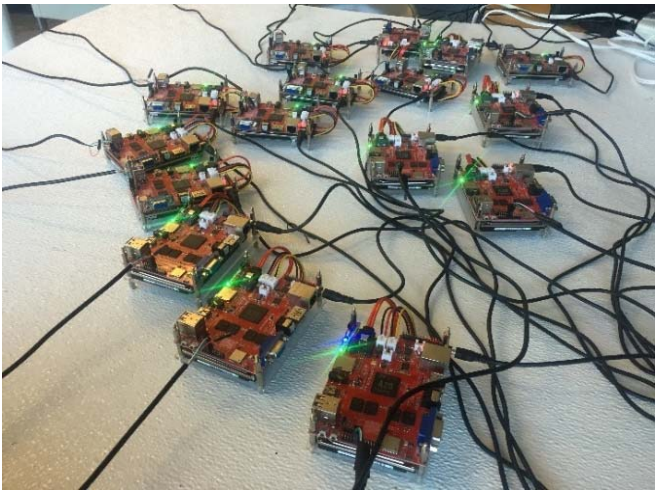

Fig. 3. 16-nodes of the Cubietruck setup

### B. 16-nodes of the Cubietruck Setup

Cubietruck is the third generation of Cubieboard. Cubieboard is a small credit-card-sized single-board computer produced by the company Cubietech. All 16 Cubietrucks used in this simulation have the Ubuntu 12 operating system installed. The code for the simulation is written in Python. Each Cubietruck will receive the data needed from the message bus. The simulation is then run and output is produced and sent to the message bus. All the connections between the Cubietrucks and the desktop computer that hosts VOLTTRON[TM] are connected by mini-USB cables. In order to extend the host machine's number of USB ports, two USB hubs are used. Fig. 3 shows an actual setup of a cluster of Cubietrucks. The parameters that characterize DGs and the consumers are selected randomly.

### C. Implementation

This section briefly lists the implementation steps of this demo system.

- VOLTTRON[TM] is installed on a Windows laptop by following the steps in the user guide [15]. The system hosting VOLTTRON[TM] is a Linux Virtual Machine (VM) run by the software Virtual Box.

- Porotype agents for the system are developed and tested on the Windows laptop.

- The formal VOLTTRON[TM] platform is then installed on a Linux desktop, which serves as the host machine.

- Serial communication between the host desktop computer and the Cubietrucks is achieved by changing the system parameters and running a Python test script. The connections between the Cubietrucks are UART to USB lines, where the UART ends are connected to the Cubietrucks.

- For the purpose of the demo, the simulation is conducted in Matlab Simulink and translated into multiple Python scripts. These Python scripts are then run in the 16-node Cubietruck setup.

- Final debugging is done with the host machine and the Cubietrucks. The heartbeat frequency is tuned to an optimal level to ensure the VOLTTRON[TM] platform provides a timely response.

### D. Results

The results are shown in Table I. The DGs are labeled as G1, G2, ..., G6 and the consumers are labeled as L1, L2, ..., L10. Nodes DG3, DG5 and L4's corresponding power outputs and demand are zero. The simulation results were validated by a centralized approach benchmark. After 42 iterations, the system reached the converging point. The total generation matched the total load, with both being around 421 kW. The incremental cost converged to 7.371 $/kWh. This rate of convergence is considered a fast one.

## VI. CONCLUSION

This study was inspired by the new challenges posed by the future smart grid and the trend of internet of things (IOT) technology, emphasizing the necessity of interconnection and communication between the devices and assets in the grid. We discussed one possible solution to these challenges, which uses an open-source agent-based platform, VOLTTRON$^{TM}$. In this paper, we introduced the functionality and typical applications of VOLTTRON$^{TM}$. First, we gave a brief overview of VOLTTRON$^{TM}$ and explained how it will be an essential tool in the future smart grid. Second, we summarized the existing projects related to VOLTTRON$^{TM}$. We also discussed the advantages of using VOLTTRON$^{TM}$ in different aspects of the smart grid. Third, we compared VOLTTRON$^{TM}$ with other similar types of software platforms and discussed their advantages and disadvantages compared with each other. Finally, we gave a brief introduction to the consensus-based distributed algorithm in [11], and provided a detailed case study of the implementation of the algorithm using VOLTTRON$^{TM}$.

REFERENCES

[1] J. Haack, B. Akyol, N. Tenney, B. Carpenter, R. Pratt, and T. Carroll," VOLTTRON™: An Agent Platform for Integrating Electric Vehicles and Smart Grid", Connected Vehicles and Expo (ICCVE), 2013 International Conference on. IEEE, 2013.

[2] S. Katipamula, J. Haack, B. Akyol, G. Hernandez, and J. Hagerman, "VOLTTRON: An open-source software platform of the future", IEEE Electrification Magazine 4.4 (2016): 15-22.

[3] A. Kantamneni, L. E. Brown, G. Parker, and W. W. Weaver," Survey of multi-agent systems for microgrid control", Engineering applications of artificial intelligence 45 (2015): 192-203.

[4] V. Chinde, A. Kohl, Z. Jiang, A. Kelkar, and S. Sarkar, "A VOLTTRONTM based implementation of Supervisory Control using Generalized Gossip for Building Energy Systems", International High Performance Buildings Conference, 2016.

[5] W. Khamphanchai, A. Saha, K. Rathinavel, M. Kuzlu, M. Pipattanasomporn, S. Rahman, B. Akyol, and J. Haack, "Conceptual architecture of building energy management open source software (BEMOSS™)", the 5th IEEE PES Intelligent Smart Grid Technologies (ISGT) European Conference Istanbul, Turkey, October 12-15, 2014.

[6] F.L. Bellifemine, G. Caire, and D. Greenwood," Developing Multi-Agent Systems with JADE", Vol. 7. John Wiley & Sons, 2007.

[7] Java Agent Development Framework (JADE), [Online]. Available: http://jade.tilab.com/

[8] F. Y. S. Eddy, "A Multi Agent System Based Control Scheme for Optimization Of Microgrid Operation", Diss. 2016.

[9] H. S. Nwana, D. T. Ndumu, L.n C. Lee and J. C. Collis,"ZEUS: A Toolkit for Building Distributed Multi-Agent Systems", Proceedings of the third annual conference on Autonomous Agents. ACM, 1999.

[10] C. Palmer, P. Lazik, M. Buevich, J. Gao, M. Berges, and A. Rowe,"Mortar.io: Open Source Building Automation System", BuildSys - ACM Int. Conf. on Embedded Systems for Energy-Efficient Built Environments , pp 204 – 205 , 2014.

[11] H. Pourbabak, T. Chen, and W. Su, "Consensus-based Distributed Control for Economic Operation of Distribution Grid with Multiple Consumers and Prosumers", 2016 IEEE Power and Energy Society General Meeting, Boston, MA, U.S.A. July 17-21, 2016.

[12] R. Olfati-Saber, J. A. Fax, and R. M. Murray, "Consensus and Cooperation in Networked Multi-Agent Systems",  Proceedings of the IEEE, vol. 95, pp. 215–233, jan 2007.

[13] R. Saber and R. Murray,"Consensus protocols for networks of dynamic agents", in Proceedings of the 2003 American Control Conference, 2003., vol. 2, pp. 951–956, IEEE, 2003.

[14] F. L. Lewis, H. Zhang, K. Hengster-Movric, and A. Das, "Cooperative Control of Multi-Agent Systems", vol. 1542 of Communications and Control Engineering, London: Springer London, 2014.

[15] VOLTTRON 3.0 User Guide, [Online]. Available: https://docs.google.com/document/d/1A7NBMGoh6Fphlf9VQW_VA9LXXSgP58Ctzz2ZqoOmjd4/edit?pref=2&pli=1

[16] H. Pourbabak, T. Chen, and B. Zhang, and W. Su, "Control and Energy Management Systems in Microgrids", Clean Energy Microgrids, Edited by Shin'ya Obara, The Institution of Engineering and Technology (IET), 2017.